\documentclass[%
 aip,
rsi,%
 amsmath,amssymb,
 reprint,%
]{revtex4-1}

\usepackage[dvipsnames,table]{xcolor}
\usepackage{xcolor}
\usepackage{graphicx}
\usepackage{dcolumn}
\usepackage{bm}
\usepackage{nicefrac}
\newcommand{\n}[1]{\mathbf{#1}}

\begin{document}

\preprint{AIP/123-QED}

\title[Rodr\'{i}guez et al.]{Solving the problem of overdetermination of quasisymmetric equilbrium solutions by near-axis expansions. II. Circular axis stellarators}

\author{E. Rodr\'{i}guez}
 \altaffiliation[Email: ]{eduardor@princeton.edu}
 \affiliation{ 
Department of Astrophysical Sciences, Princeton University, Princeton, NJ, 08543
}
\affiliation{%
Princeton Plasma Physics Laboratory, Princeton, NJ, 08540
}%

\author{A. Bhattacharjee}
 \altaffiliation[Email: ]{amitava@princeton.edu}
 \affiliation{ 
Department of Astrophysical Sciences, Princeton University, Princeton, NJ, 08543
}
\affiliation{%
Princeton Plasma Physics Laboratory, Princeton, NJ, 08540
}%

\date{\today}

\begin{abstract}
We apply the near-axis expansion method for quasisymmetric magnetic fields with anisotropic pressure (developed in the companion paper, Part I)\cite{rodr2020} to construct numerical solutions to circular axis stellarators. The solutions are found to second order in the distance from the axis, not possible in the standard Garren-Boozer construction [D. A. Garren and A. H. Boozer, Physics of Fluids B: Plasma Physics, 2822 (1991)], which assumes magnetostatic equilibria with isotropic pressure.\cite{garrenboozer1991b} In the limit of zero anisotropy, it is shown that a subset of coefficients can be chosen to avoid the overdetermination problem. 
\end{abstract}

\maketitle

\section{Introduction:}\label{sec:intro}

\textit{Quasisymmetric} magnetic fields as a basis for constructing stellarators are a natural starting point to build good plasma confinement devices. \textit{Quasisymmetry} in its \textit{weak} sense, despite not being a continuous symmetry, confers magnetized charged particles with an approximate conserved canonical momentum. This keeps particles close to magnetic flux surfaces, to which magnetic field lines $\n{B}$ are tangent. Such configurations have been explored in many instances in the literature.\cite{nuhren1988,garebidian1996,Henneberg2019} \par
However attractive, it is believed that fields in magnetostatic (MS) equilibrium with isotropic pressure that are \textit{quasisymmetric} (QS) in a global sense are likely not to exist. This point of view found strong support in the important work of Garren and Boozer\cite{garrenboozer1991b}. Constructing solutions by expansion around the magnetic axis, they showed that a QS magnetic field in MS equilibrium is forced to satisfy a set of overdetermined equations, that is, the equations must satisfy a larger number of constraints than the allowed degrees of freedom. \par

In Part I of this sequence of papers\cite{rodr2020}, and following the insight in [\onlinecite{burby2019}] and [\onlinecite{rodriguez2020}], the near-axis expansion procedure was decoupled from the requirements of MS equilibria with scalar pressure. Doing so allowed us to introduce a more general form of force balance, one that admits anisotropic pressure. In this paper we shall apply this new approach to the particular case of a circular magnetic axis to explicitly construct solutions through second order. The purpose of choosing this example is two-fold. First, it enables us to show the new construction at work to second order in as simple a case as possible, for which the problem of overdetermination occurs for MS equilibria with scalar pressure [\onlinecite{garrenboozer1991b}]. The so called \textit{looped} differential equations of the construction obtained in Part I, appear for the first time at this order. Second, in the absence of a rigorous proof of the existence of solutions of the looped equations, an explicit numerical solution through second order would be a practical demonstration of how to construct such solutions.  \par
In addition to the relevance of such solutions for a restricted class of stellarators, these are also of interest for "three-dimensional" tokamaks (i.e., classical axisymmetric tokamaks plus symmetry-breaking perturbations).

The following is the plan for this paper. In Section II, we start with a practical step-by-step application of the first two orders of the general construction in Part I, applied to a circular axis. Special attention is paid to the differential equations involved and their solution. In Section III, we follow up with a numerical example. Finally, we explore how the isotropic limit may be achieved, and close up with conclusions.

\section{Construction of solution}
\begin{figure}
   \centering
   \includegraphics[width=0.3\textwidth]{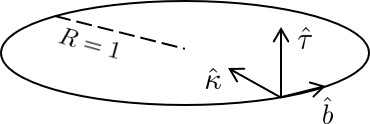}
    \caption{\textbf{Circular magnetic axis.} Depiction of the circular magnetic axis for the stellarator construction, showing the well-defined Frenet basis. The unit vectors are: $\hat{b}$ tangent, $\hat{\kappa}$ curvature and $\hat{\tau}$ binormal.}
    \label{fig:circAxis}
\end{figure}

Let us consider the simple case of a \textit{circular axis}. Such an axis is fully characterised by a constant curvature ($\kappa=1$) and a vanishing torsion ($\tau=0$). The radius of the axis is taken to have unit length, with all other length measures taken relative to it. As the curvature vanishes nowhere, this guarantees the existence of a well-defined Frenet set of vectors everywhere along the axis (see Fig. \ref{fig:circAxis}). \par
The solution off axis will be constructed order-by-order following Part I. The labels used to refer to specific equations follow the convention introduced in Part I as well, which we shall not reproduce here. The main definitions of functions and relevant expansions inherited from Part I are included in Appendix A for completeness.

\subsection{Expansion to first order}
Start off with $J^1$,
\begin{equation*}
    B_{\alpha 0}=\frac{\mathrm{d}l}{\mathrm{d}\phi}\frac{1}{\sqrt{B_0}},
\end{equation*}
where by construction $B_0$ and $B_{\alpha 0}$ are constant. Thus, it follows that $\mathrm{d}l/\mathrm{d}\phi$ is also a constant. Now, if we integrate it over a whole toroidal loop, it follows that $\mathrm{d}l/\mathrm{d}\phi=1$. (Thus, in this case, $\phi$ corresponds to the standard cylindrical angle.) \par
Let us now consider the functions $X_1$ and $Y_1$, which together describe the shape of flux surfaces to leading order. The former is obtained from the next order Jacobian equation, (Eqs. (14) and (15) in Part I)
\begin{align*}
    X_{11}^C=-\frac{B_{\alpha 0}^2B_{11}^C}{2}=\eta, \\
    X_{11}^S=0,
\end{align*}
which are constant and correspond to a choice of coordinates that make $1/B^2$ have only a cosine component to leading order. \par
The leading order \textit{co(ntra)variant} equations describe the function $Y_1$, with $Z_1=0=B_{\theta 0}=B_{\theta 1}$,. From $C_b^0$, (Eq. (23) in Part I), we obtain
\begin{equation*}
    Y_{11}^S=-\frac{4\sqrt{B_0}}{B_{11}^CB_{\alpha0}^2}=\frac{2\sqrt{B_0}}{\eta},
\end{equation*}
which, like $X_{11}^C$, is again a constant. For the sine component of $Y_1$, $C_\perp^1$ yields a first-order ordinary differential equation (ODE), which using the customary definition $Y_{11}^C=Y_{11}^S\sigma$, reads (Eq. (26) in Part I)
\begin{equation}
    \frac{\mathrm{d}\sigma}{\mathrm{d}\phi}+\Bar{\iota}_0\sigma^2+\frac{\Bar{\iota}_0}{4B_0}(\eta^4+4B_0)-\frac{\eta^2}{2\sqrt{B_0}}B_{\theta 20}^C=0. \label{eq:diffEqSigma}
\end{equation}
The Riccati equation for $\sigma$ must be completed with an appropriate boundary condition. In this case, we impose $\sigma(0)=\sigma(2\pi)$ to guarantee periodicity. \par
Before discussing this equation in some more detail, we note that $\sigma$ is not the only unknown in Eq. (\ref{eq:diffEqSigma}). We have no knowledge about the function $B_{\theta20}$, which although second order should be found now. Using equation II$^2$ (Eq. (44) in Part I), we obtain
\begin{align}
    &B_{\theta20}^C(1-\Delta_0)=\nonumber\\
    &=\Bar{B}_{\theta20}-B_{\alpha0}\frac{\Bar{\iota}_0\eta^2}{2}\sum_{n=0}^\infty\frac{1}{\Bar{\iota}_0^2-n^2}\left(\Bar{\Delta}_{0n}^S\sin n\phi+\Bar{\Delta}_{0n}^C\cos n\phi\right), \label{eq:Btheta20}
\end{align}
where $\Delta_{0n}^{S/C}$ represent coefficients (sine and cosine) of the Fourier series in $\phi$ of the anisotropy on axis, $\Delta_0$. The constant $\Bar{B}_{\theta20}$ is a free coefficient physically related to the plasma current density at the magnetic axis. With the closed form for $B_{\theta20}$ in (\ref{eq:Btheta20}), the non-linear ODE for $\sigma$ may now be solved. For a thorough analysis of equation (\ref{eq:diffEqSigma}), we refer the reader to, for example, [\onlinecite{landreman2019}]. Here, we restrict the discussion to the main aspects of the equation without proof. \par
Equation (\ref{eq:diffEqSigma}) may be formulated as an initial-value problem for some $\sigma(0)$. Then one can prove\cite{landreman2019} that for each $\sigma(0)$, there exists a value for the rotational transform $\Bar{\iota}_0$ on axis such that the solution to the equation is periodic in $\phi$. Numerically, the initial-value problem in $\sigma$ can be solved, for example, by a standard Runge-Kutta fourth- order scheme for every pair of parameters $(\sigma(0),\Bar{\iota}_0)$. As a result of this integration, one can construct the quantity $\sigma(2\pi)-\sigma(0)$. To obtain periodic solutions, the next step is to use a standard Newton method to obtain a zero of this expression.\cite{landreman2019} \par
Given the freedom there exists in the choice of $\sigma(0)$, it is important to understand its physical meaning. By definition, $\sigma$ forms part of the function $Y_1$, and thus describes some aspect of the shape of magnetic flux surfaces to first order. The loci of such flux surfaces are given by,
\begin{equation*}
    \n{x}-\n{r}_0=\epsilon\eta\cos \chi\hat{\kappa}+\frac{2\epsilon}{\eta B_{\alpha0}}(\sin\chi+\sigma\cos\chi)\hat{\tau}.
\end{equation*}
\begin{figure}
    \centering
    \includegraphics[width=0.25\textwidth]{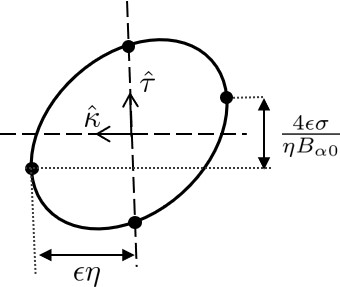}
     \caption{\textbf{Leading order flux surface.} Representation of the flux surfaces to leading order around the magnetic axis. This shows, in particular, the physical meaning of the function $\sigma$, and parameters such as $\eta$.} 
     \label{fig:ellipFlux}
 \end{figure}
With the help of Fig. \ref{fig:ellipFlux}, which represents a poloidal cross section of the stellarator, $\sigma$ may be seen to affect both the height and tilt of the elliptical cross-section of a magnetic flux surface. Thus, a large value of $\sigma$ produces a predominant elongation in the binormal direction, with $\sigma=0$ representing alignment with the Frenet axes. \par
On a slightly different note, additional important information about the field may be obtained by following the point of largest $B$ around the torus, represented by the leftmost black dot in Fig. \ref{fig:ellipFlux}.\cite{landreman2018a} Going around the torus once, the vector $\n{x}-\n{r}_0$ remains within $\pi/2~$rad measured with respect to the curvature vector. This forces the contour to have the same helical behaviour as the curvature of the axis. In our case, the field will thus be necessarily quasiaxisymmetric. \par
Before moving on to the next order, we make some remarks on the role of $\bar{B}_{\theta 20}$. Its role and relation to other coefficients is clear once the MS limit ($\Delta_0\rightarrow0$) is considered. In this limit $B_{\theta20}=\bar{B}_{\theta20}$, $\sigma=\sigma(0)$, and the rotational transform becomes
\begin{equation}
    \Bar{\iota}_0=\frac{\eta^2}{2\sqrt{B_0}}B_{\theta20}\left(1+\frac{\eta^4}{4B_0}+\sigma(0)^2\right)^{-1}. \label{eq:estimateIota}
\end{equation}
This shows the importance of $B_{\theta20}$ both in providing a $\phi$-dependent shaping of flux surfaces, as well as setting the value of the rotational transform on axis. The former implies that for moderate anisotropy flux surfaces will not be far from being axisymmetric to leading order. The latter indicates that the rotational transform on axis originates from the on-axis current, which makes the stellarator a physically unattractive candidate (more like a tokamak). For moderate anisotropy, the Newton finder described before will yield values close to (\ref{eq:estimateIota}). \par
To complete the first-order construction, expressions for the pressure and anisotropy through first order are needed. These are precisely of the same form as presented in Part I, and thus for brevity we do not repeat them here.

\subsection{Expansion to second order}
The first-order expansion described above is very similar to the MS case worked out previously, for example, in [\onlinecite{landreman2018a}]. However, it is at second order that our approach produces qualitatively different results. Whereas in the standard approach the second-order equation appears to be overdetermined\cite{garrenboozer1991b,landreman2018a}, in the new approach we avoid the problem of overdetermination and are able to construct a solution through second order explicitly. \par
\begin{figure}
    \includegraphics[width=0.5\textwidth]{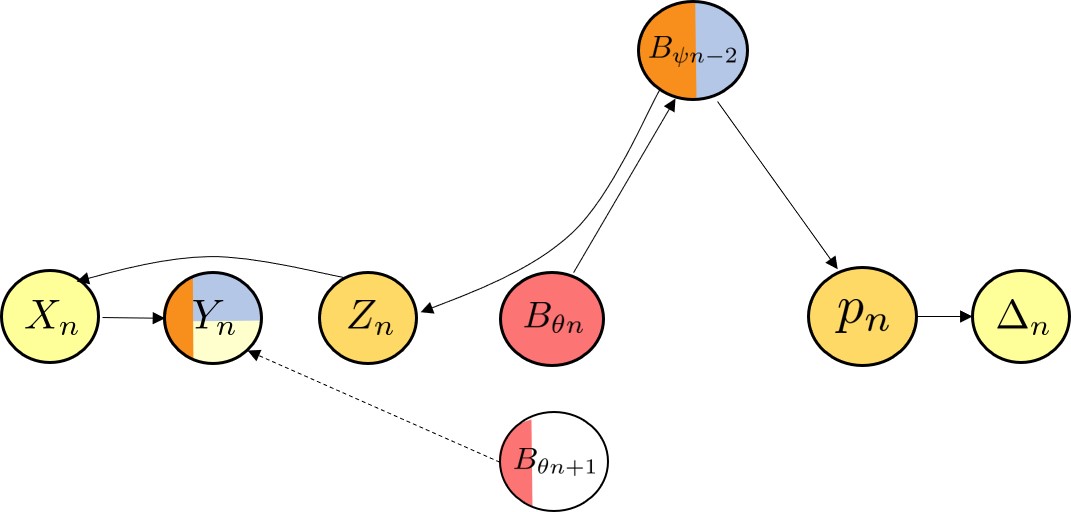}
    \caption{\textbf{Consistent order of functions.} Order of functions that ought to be simultaneously solved for. The color code represents the order in which the functions are solved for: red - first to be solved using the \textit{looped} form of II, dark orange, bright orange, yellow and pale yellow. The blue color represents the introduction of free functions at even orders that are to be solved self consistently. The dashed arrow and reddish colored part of $B_{\theta n+1}$ represent that for odd orders, the 0 harmonic term of the next order is to be solved simultaneously as well. The arrows show the mutual dependencies, originating from $B_{\theta n}$.}
    \label{fig:ConstructNAAnis}
\end{figure}
As pointed out earlier in this paper as well as in Part I, a noteworthy feature that appears at second order is the set of two self-consistent, so-called, \textit{loop} equations. The unknowns in these equations are, as will be made explicit later, the functions $Y_{20}$ and $B_{\psi 0}$. To obtain the \textit{looped} equations explicitly in these two unknowns, it is essential to know the form of the rest of the second order functions, especially given the degree of mutual dependence, as illustrated in Fig. \ref{fig:ConstructNAAnis} (diagram from Part I, reproduced again here). \par
So let us start with $Z_2$. The only harmonic of $Z_2$ that involves $B_{\psi0}$ reads, (Eqs. (14) and (24) in Part I)
\begin{gather*}
    Z_{2,0}=\frac{B_{\psi0}}{B_{\alpha0}}+\Tilde{Z}_{20},
\end{gather*}
where
\begin{equation*}
    \tilde{Z}_{20}=-B_0\frac{\sigma\sigma'}{\eta^2}.
\end{equation*}
The other two components are all written in terms of expressions already known, that is,
\begin{gather*}
    Z_{2,2}^S=B_0\frac{\Bar{\iota}_0}{\eta^2}\left[\sigma^2-1+\frac{\eta^4}{4B_0}\right]-B_0\frac{\Bar{\iota}_0}{\eta^2}\sigma', \\
    Z_{2,2}^C=-B_0\frac{\sigma(2\Bar{\iota}_0+\sigma')}{\eta^2}.
\end{gather*}
Following the flow in Fig. \ref{fig:ConstructNAAnis}, we construct $X_2$ using $J^2$. Again, as it happened for $Z_2$, only one of the harmonics includes $B_\psi$, (see Appendix C in Part I)
\begin{equation*}
    X_{2,0}=\frac{B_{\psi0}'}{B_{\alpha0}}+\Tilde{X}_{2,0}
\end{equation*}
where,
\begin{align*}
    \Tilde{X}_{2,0}=\frac{1}{4}&\left[-4\frac{B_{\alpha1}}{B_{\alpha0}}+\eta^2(1+\Bar{\iota}_0^2)-2\frac{B_{20}}{B_0}+\frac{4B_0}{\eta^2}\Bar{\iota}_0^2(1+\sigma^2)+\right.,\\
    &\left.+\frac{8B_0\Bar{\iota}_0}{\eta^2}-\frac{4B_0}{\eta^2}\sigma\sigma''\right].
\end{align*}
The other two harmonics read
\begin{align*}
    X_{2,2}^C=&\frac{B_0}{4\eta^2}\left[\frac{\eta^4}{B_0}(1+\Bar{\iota}_0^2)-2\eta^2\frac{B_{22}^C}{B_0^2}+4\Bar{\iota}_0^2(\sigma^2-1)-\right.\\
    &\left.-8\Bar{\iota}_0\sigma'-4\sigma\sigma''\right],  \\
    X_{2,2}^S=&-\frac{B_0}{2\eta^2}\left[\eta^2\frac{B_{22}^S}{B_0^2}-4\Bar{\iota}_0^2\sigma-4\Bar{\iota}_0\sigma\sigma'+2\sigma''\right].
\end{align*}
\par
The last of the functions describing the flux surfaces through second order is $Y_2$. In this case, using $C_b^1$, (Eqs. (27) and (28) in Part I), we obtain
\begin{gather*}
    Y_{2,2}^C=Y_{2,0}-\frac{2B_0\sigma}{\eta^2}B_{\psi0}'+\Tilde{Y}_{2,2}^C, \\
    Y_{2,2}^S=-\frac{2B_0}{\eta^2}B_{\psi0}'+\Tilde{Y}_{2,2}^S,
\end{gather*} 
where,
\begin{gather*}
    \Tilde{Y}_{2,2}^C=\frac{2}{B_{\alpha0}\eta^2}\left[\left(X_{2,2}^C-\Tilde{X}_{2,0}\right)+X_{2,2}^S\right], \\
    \Tilde{Y}_{2,2}^S=-\frac{1}{B_{\alpha0}}-\frac{2}{B_{\alpha0}\eta^2}\left[\left(X_{2,2}^C+\Tilde{X}_{2,0}\right)-X_{2,2}^S\sigma\right].
\end{gather*}
To continue constructing functions at second-order, let us turn our view to the opposite side of Fig. \ref{fig:ConstructNAAnis}. Consider the perpendicular pressure, whose 0th harmonic is the only component explicitly dependent on $B_{\psi0}$. From III$^0$ (Eqn. (45) in Part I),
\begin{equation*}
    p_{2,0}^C=-B_{\alpha0}\left[B_{\psi0}(\Delta_0-1)\right]'+\Tilde{p}_{20}
\end{equation*}
where,
\begin{align*}
    \Tilde{p}_{2,0}=&\frac{B_{\alpha0}}{8B_0}\left[8B_0B_{\alpha1}(\Delta_0-1)+B_{\alpha0}(4B_{20}\Delta_0-\right.\\
    &\left.-2B_0\eta(4\eta\Delta_0+\Delta_{11}^C))\right].
\end{align*}
The other two harmonics read,
\begin{gather*}
    p_{2,2}^C=\frac{1}{8B_0^2}\left[4B_{22}^C\Delta_0-2B_0\eta(4\eta\Delta_0+\Delta_{11}^C)\right],\\
    p_{2,2}^S=\frac{1}{4B_0^2}\left[2B_{2,2}^S\Delta_0-B_0\eta\Delta_{11}^S\right].
\end{gather*}
Finally, we require the second-order form of the anisotropy $\Delta$, which will be constructed using Eq. I$^2$. From Eq. (62) of Part I,
\begin{equation*}
    \Delta_{2,0}=-B_0p_{2,0}+\Tilde{\Delta}_{20}+\left\langle B_0p_{2,0}-\Tilde{\Delta}_{20}\right\rangle_\phi+\Bar{\Delta}_{20},
\end{equation*}
where
\begin{align*}
    \Tilde{\Delta}_{2,0}=&\eta(2B_0p_{11}^C+\Delta_{11}^C)+\frac{B_{20}}{B_0}\Delta_0+\frac{(B_{11}^C)^2}{2B_0^2}\Delta_0-\\
    &-\frac{\Bar{\iota}_0B_{11}^C}{4B_0}\int(4B_0p_{11}^S+3\Delta_{11}^S)\mathrm{d}\phi.
\end{align*}
The constant $\Bar{\Delta}_{20}$ is a free parameter of choice, defined in a way that it represents the $\phi$-averaged value of the anisotropy through second order. \par
To obtain the other two harmonics of $\Delta$, simple harmonic oscillator (SHO) equations (see Eqs. (63) and (64) in Part I for more details and definitions) of the form 
\begin{equation*}
    (B_0p_{22}^C+\Delta_{22}^C)''+(2\Bar{\iota}_0)^2(B_0p_{22}^C+\Delta_{22}^C)-f=0,
\end{equation*}
need to be solved, where $f=2\Bar{\iota}_0\mathcal{B}-\mathcal{A}'$.  \par
Interested in a particular solution to the equation due to periodicity requirements, the problem may be solved straightforwardly in discrete Fourier space. Numerically, the procedure requires finding the discrete Fourier coefficients of $f$, applying $n$-dependent factors to them,
\begin{equation*}
    \Delta_{22}^C=-B_0p_{22}^C+\sum_{n=0}^\infty\frac{1}{(2\Bar{\iota}_0)^2-n^2}(f_n^C\cos n\phi+f_n^S\sin n\phi).
\end{equation*}
For a well-behaved $\Delta_0$, it will suffice to retain the first few Fourier harmonics for $f$. \par
This explicit construction of $\Delta_2$ has been included not only for completeness, but also because it shows that rational values of $\bar{\iota}_0$ can constitute a problem. At this order, a transform $\Bar{\iota}_0=1/2$ leads to a vanishing denominator for $n=1$, making $\Delta_{22}$ blow up. More generally, if we write $\Bar{\iota}_0=m/n$, then $\Delta_n$ is expected to diverge, except if the $m$-th harmonic of $f$ is zero. These resonances should be avoided. Some numerical examples are shown later. \par

\subsection{Construction of \textit{looped} equation II}
With all the relevant functions to order $n=2$ explicitly constructed and expressed in terms of the unknowns $B_{\psi0}$ and $Y_{20}$, it is now the time to deal with the \textit{looped} equation $\Tilde{\mathrm{II}}_\mathrm{SC}^3$. To do so we will follow the instructions detailed in Part I. \par
The algebraic details of this construction are left to Appendix B. Here the main steps in the procedure are sketched. First we should stablish what the \textit{looped} equation $\Tilde{\mathrm{II}}^3$ alludes to. The equation is a modified version of Eq. II$^3$ in which all the $n=3$ order functions have been eliminated in favour of $Y_{20}$ and $B_{\psi0}$. \par
To obtain it, we start by writing down II$^3$ explicitly, i.e. two differential equations in $B_{\theta31}$. These equations depend on $p_{3,1}$ explicitly, which should be eliminated. To do so, we need III$^1$, which will in turn involve $B_{\psi1,1}$. Finally, using $C_\perp^2$, the expression remains free of $B_{\theta3,1}$, $p_{3,1}$ and $B_{\psi11}$ altogether. The system of coupled differential equations may then be explicitly written as,
\begin{align}
    C_{B1}^1 B_{\psi0}'+C_{B2}^1 B_{\psi0}''+C_{Y0}^1& Y_{20}+C_{Y1}^1 Y_{20}'+\nonumber\\
    &+C_{Y2}^2 Y_{20}''+C_0^1=0, \label{eq:circLoopIIa}
\end{align}
and
\begin{align}
    C_{B1}^2 B_{\psi0}'+C_{B2}^2 B_{\psi0}''+C_{B3}^2& B_{\psi0}'''+C_{Y0}^2 Y_{20}+\nonumber\\
    &+C_{Y1}^2 Y_{20}'+C_0^2=0.\label{eq:circLoopIIb}
\end{align}
The precise form of the coefficients $C$ may be found in Appendix B, Eqs. (\ref{eq:circLoopIIafull}) and (\ref{eq:circLoopIIbfull}). It is important to note that these factors are independent of the unknowns $B_{\psi0}'$ and $Y_{20}$. \par
Equations (\ref{eq:circLoopIIa}) and (\ref{eq:circLoopIIb}) constitute the \textit{looped} equations $\Tilde{\mathrm{II}}_\mathrm{SC}^3$: a system of coupled linear differential equations for $B_{\psi0}'$ and $Y_{20}$. The highest derivative in each of the equations is third order. In fact, the equation associated with the sine component, Eq. (\ref{eq:circLoopIIa}), has as leading derivative terms $B_{\psi0}''$ and $Y_{20}''$. That associated with the cosine component, Eqn. (\ref{eq:circLoopIIb}), involves $B_{\psi0}'''$ and $Y_{20}'$. \par
With all the coefficients in this equation known, a solution needs to be found numerically. Because we are interested in finding periodic solutions, and the coupled set of differential equations is linear, it is convenient to represent all the relevant functions in a periodic basis in $\phi$. This representation will, by construction, guarantee the periodicity of the unknowns of the problem, which we will refer to in shorthand now as $\lambda$. Writing
\begin{equation*}
    \lambda=\sum_{n=-n_\lambda}^{n_\lambda} \lambda_ne^{in\phi},
\end{equation*}
the ordinary differential equations are now transformed into algebraic systems of equations in $\lambda_n$. To successfully complete this transformation though, the action of two main operations need to be specified. First, we need to know how derivatives translate in this new representation. This is straightforward since $\lambda'=\sum in\lambda_n\exp(in\phi)$.  \par
The second operation of importance is the product of the coefficients $C$ with the unknown functions $\lambda$ in this Fourier basis. Because we are dealing with the problem numerically, we will approximate the function $\lambda$ by a finite sum including $n_\lambda$ harmonics. Similarly, let the coefficients in equations (\ref{eq:circLoopIIa}) and (\ref{eq:circLoopIIb}) be expressed as finite Fourier series limited to $n_C<n_\lambda$ harmonics, with coefficients $C_n$. The product then reads, 
\begin{equation*}
    C\lambda=\sum_{l=-(n_\lambda+n_C)}^{n_\lambda+n_C}\sum_{n=-n_\lambda}^{n_\lambda}\lambda_nC_{l-n}e^{il\phi}.
\end{equation*}
With this in mind, from each of equations (\ref{eq:circLoopIIa}) and (\ref{eq:circLoopIIb}) we will have a total of $2(n_\lambda+n_C)+1$ algebraic equations, one for each of the harmonics $\exp(il\phi)$. The Fourier coefficients for $Y_{20}$ and $B_{\psi0}'$ constitute a total of $2(2n_\lambda+1)$ unknowns, which would make the algebraic system overconstrained. An appropriate truncation is needed for the numerical problem to be well-posed. Assuming higher harmonics to become increasingly more negligible, only a small truncation error is introduced by keeping just the lower $n_\lambda$ harmonics equations. \par
The linear system of algebraic equations may be succinctly put in matrix form, defining the matrices
\begin{gather*}
    A_{l,n}=(C_{B1}^1)_{l-n}+in(C_{B2}^1)_{l-n} \\
    B_{l,n}=(C_{Y0}^1)_{l-n}+in(C_{Y1}^1)_{l-n}+l^2\delta_{nl} \\
    C_{l,n}=(C_{B1}^2)_{l-n}+in(C_{B2}^2)_{l-n}+l^2\delta_{nl} \\
    D_{l,n}=(C_{Y0}^2)_{l-n}+in(C_{Y1}^2)_{l-n},   
\end{gather*}
where subscripts outside brackets label harmonics of the coefficient inside (and are taken to vanish if $|l-n|>n_C$), and $\delta_{nl}$ is the Kronecker delta function. These matrices are dominantly diagonal, with $n_C$ regulating the strength of the coupling between different row/columns. The system of equations reads,
\begin{equation*}
    \begin{bmatrix}
        & A & & & B &  \\
        & C & & & D & 
    \end{bmatrix}
    \lambda=
    -\begin{bmatrix}
        C_0^1 \\
        \vdots \\
        C_0^2
    \end{bmatrix},
\end{equation*}
where $\lambda$ is a vector including the harmonics of $B_{\psi0}'$ and $Y_{20}$, and the right-hand-side contains the harmonics of the inhomogeneous terms. \par
A periodic solution for $B_{\psi0}'$ and $Y_{20}$ is found by numerically inverting the coefficient matrix (generally non-singular), for which efficient algorithms exist (especially if the matrix is sparse). Solutions found this way may be then verified by feeding the values of the solution at $\phi=0$ to a Runge-Kutta solver.\par
This does not however conclude the solution finding, because for the solution $B_{\psi0}'$ to be valid, its integral over $\phi$ in the domain $[0,2\pi)$ (or equivalently the 0th harmonic) must vanish. (Otherwise, the function $B_{\psi0}$ would be multi-valued, which would be unphysical). In order to satisfy this additional property, it is necessary to make specific choices for parameters such as $\bar{B}_{\theta20}$, $B_{22}$, $\bar{\Delta}_{20}$, etc. At second order, there is only one new free parameter $\Bar{\Delta}_{20}$. By optimising this single parameter a solution that satisfies the requirements is sought. While we provide here no proof that this procedure always yields a solution, we are able to provide explicit numerical solutions and an interpretation of the solutions in the isotropic limit that suggest this procedure can be implemented with success. \par
\begin{figure}[]
    \centering
    \includegraphics[width=0.45\textwidth]{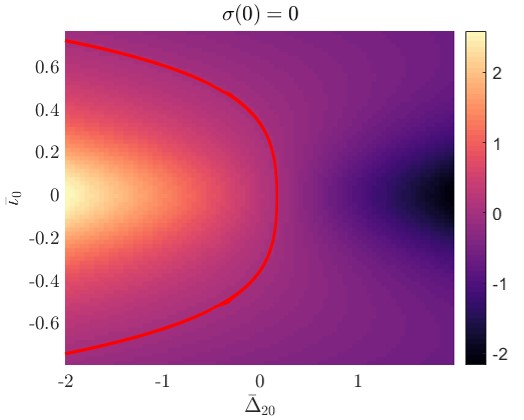}
     \caption{\textbf{Optimisation space for correct solution.} The figure shows the value of $\int_0^{2\pi}B_{\psi0}'\mathrm{d}\phi$ in a space spanned by parameters $\Bar{\iota}_0$ and $\Bar{\Delta}_{20}$. The red line represents the loci of valid solutions. The figure was created for field coefficients as specified in Section III, but with $\Delta_0=0$.}
     \label{fig:paramSpcS0D0}
\end{figure}

An illustrative example of this optimisation in $\bar{\Delta}_{20}$ for a number of values of $\bar{\iota}_0$ is shown in Fig. \ref{fig:paramSpcS0D0}. The color represents the value of $\int B_{\psi0}'\mathrm{d}\phi$, with the red line representing the set of valid solutions through second order.


\section{A numerical example}
Let us now construct a numerical example through second order. Take the on-axis anisotropy to be described by two harmonics of the form $\Delta_0=-0.1\cos\phi-0.1\cos2\phi$ for example, with $p_\perp=1-\Delta_0$. For the magnetic field magnitude, let $1/B^2=1-1.8\epsilon\cos\theta+0.01\epsilon^2(1+\cos2\theta+\sin2\theta)$ and $B_\alpha=1+0.1\epsilon^2$. Taking $\sigma(0)=0$ and $\Bar{\iota}_0=0.53$, the problem specification is complete. \par
\begin{figure}[]
    \centering
    \includegraphics[width=0.5\textwidth]{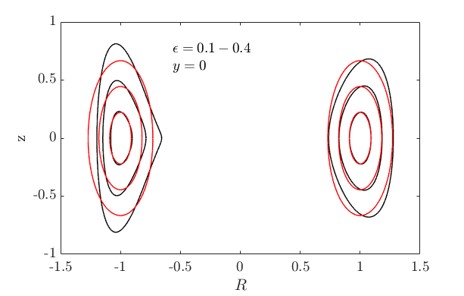}
     \caption{\textbf{Cross section of flux surfaces.} The figure shows the cross- sections of flux surfaces at $\epsilon=0.1,~0.2,~0.3$ at $\phi=0,\pi$. The red contour shows the surfaces to leading order, while the black contours include the second order corrections.}
     \label{fig:stellCrossSec}
\end{figure}

First, let us construct the magnetic flux surfaces, and show the Poincar\'{e} plot corresponding to a cut through $\phi=0,~\pi$ (the geometric cylindrical coordinate). The contours are shown in Fig. \ref{fig:stellCrossSec}, where a few surfaces corresponding to different values of $\epsilon$ are plotted. Two aspects illustrated by this contours are of importance. First, the solution lacks axisymmetry, but it is quasisymmetric by construction (within numerical error). Second, flux surfaces do not intersect up to $\epsilon\sim0.4$, above which a solution cannot exist. This limits how small the aspect ratio of the stellarator can be made. Surfaces further away from the axis are also correct only approximately to second order. As qualitative guidance for the significance of the corrections, we show a comparison of the first- and second-order surfaces. To obtain more precise results, it would be necessary to compute $\epsilon^3$ corrections, but these lie outside the scope of the present paper. It is interesting to note that the second-order correction brings negative triangularity into the problem. \par
\begin{figure*}[]
    \hspace*{-1cm}
    \includegraphics[width=1\textwidth]{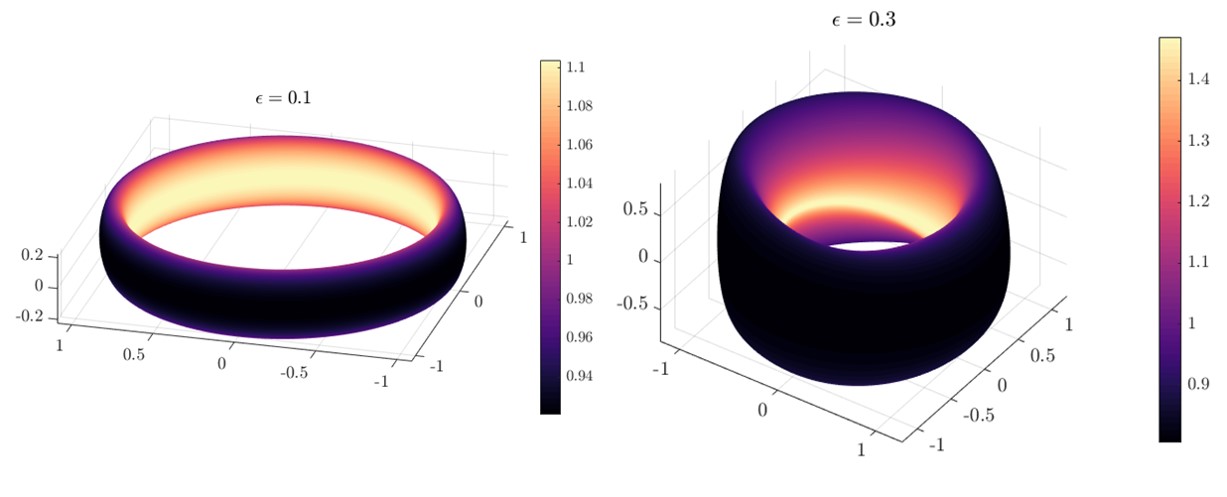}
     \caption{\textbf{Magnetic field at flux surfaces for circular axis stellarator.} The figures show the magnetic field magnitude on two magnetic flux surfaces at $\epsilon=0.1$ and $\epsilon=0.3$, for the parameters specified in this section.}
     \label{fig:stellBmag}
\end{figure*}

\begin{figure*}[]
    \hspace*{-1cm}
    \includegraphics[width=1\textwidth]{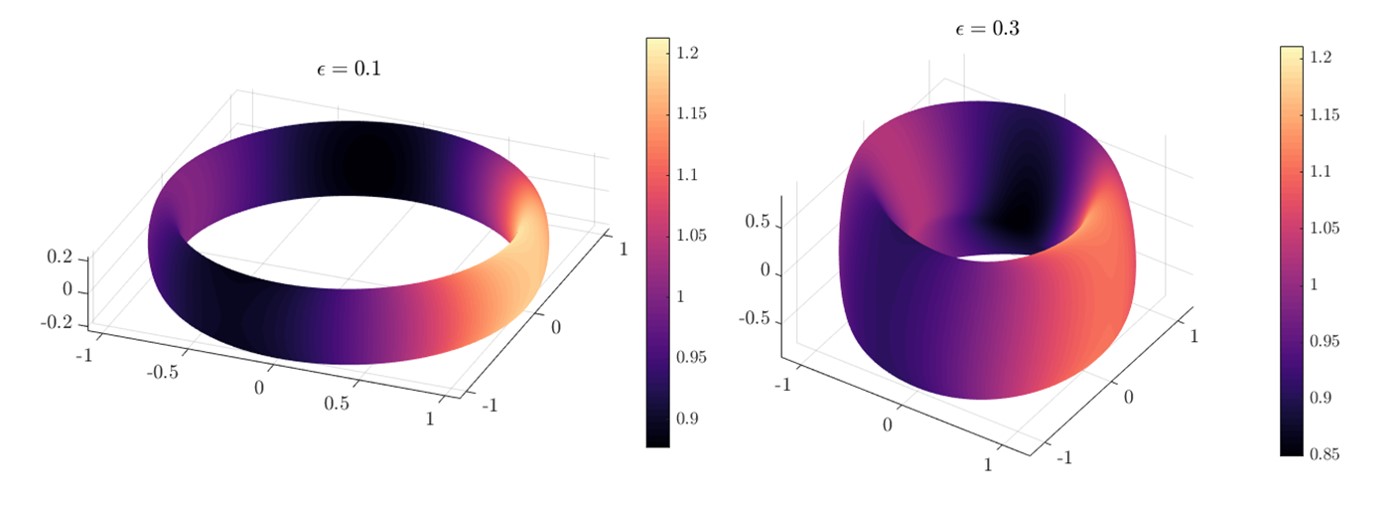}
     \caption{\textbf{Perpendicular pressure, $p_\perp$, at flux surfaces for circular axis stellarator.} The figures show the perpendicular pressure on two magnetic flux surfaces at $\epsilon=0.1$ and $\epsilon=0.3$, for the parameters specified in this section.}
     \label{fig:stellPress}
\end{figure*}

\begin{figure*}[]
    \hspace*{-1cm}
    \includegraphics[width=1\textwidth]{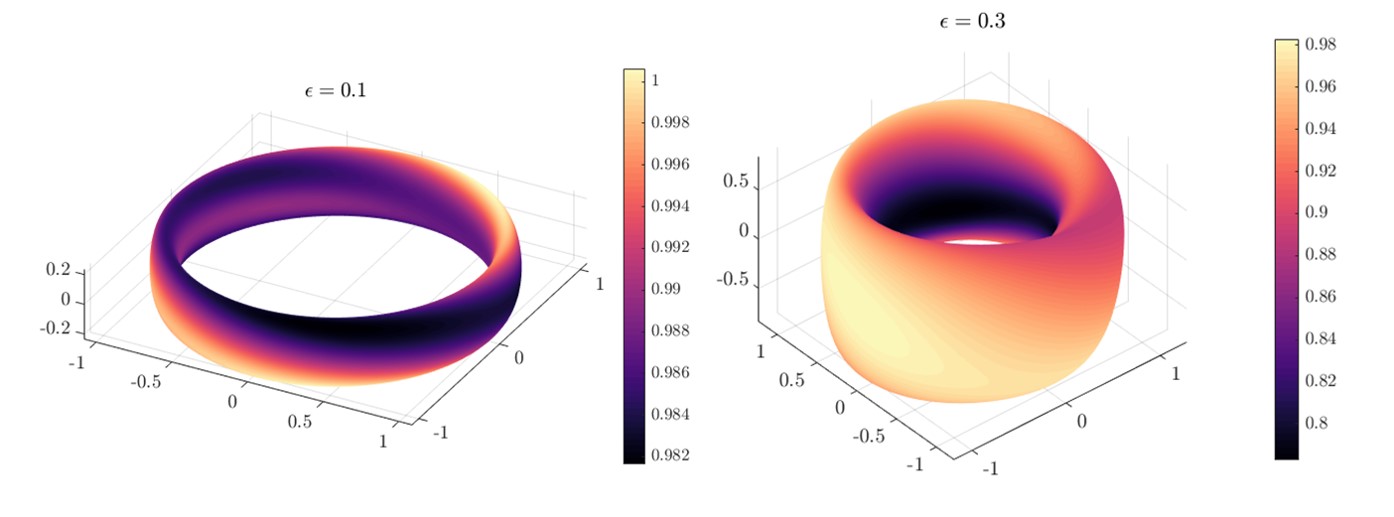}
     \caption{\textbf{Parallel pressure, $p_\parallel$, at flux surfaces for circular axis stellarator.} The figures show the parallel pressure on two magnetic flux surfaces at $\epsilon=0.1$ and $\epsilon=0.3$, for the parameters specified in this section.}
     \label{fig:stellParPress}
\end{figure*}

To complete the solution we need to explore the fully three-dimensional structure of the stellarator. It is also important to check that the functions $p_\perp$, $p_\parallel$ and $B$ are physically realizable. Figure \ref{fig:stellBmag} shows the value of the quasiaxisymmetric magnetic field on two magnetic flux surfaces, while the perpendicular and parallel pressures are shown in Figs. \ref{fig:stellPress} and \ref{fig:stellParPress} respectively. \par
The departure of the equilibrium from isotropy is evident, both from the misalignment of flux surfaces and constant pressure surfaces, as well as $p_\parallel\neq p_\perp$. \par
\begin{figure*}[]
    \hspace*{-1cm}
    \includegraphics[width=1\textwidth]{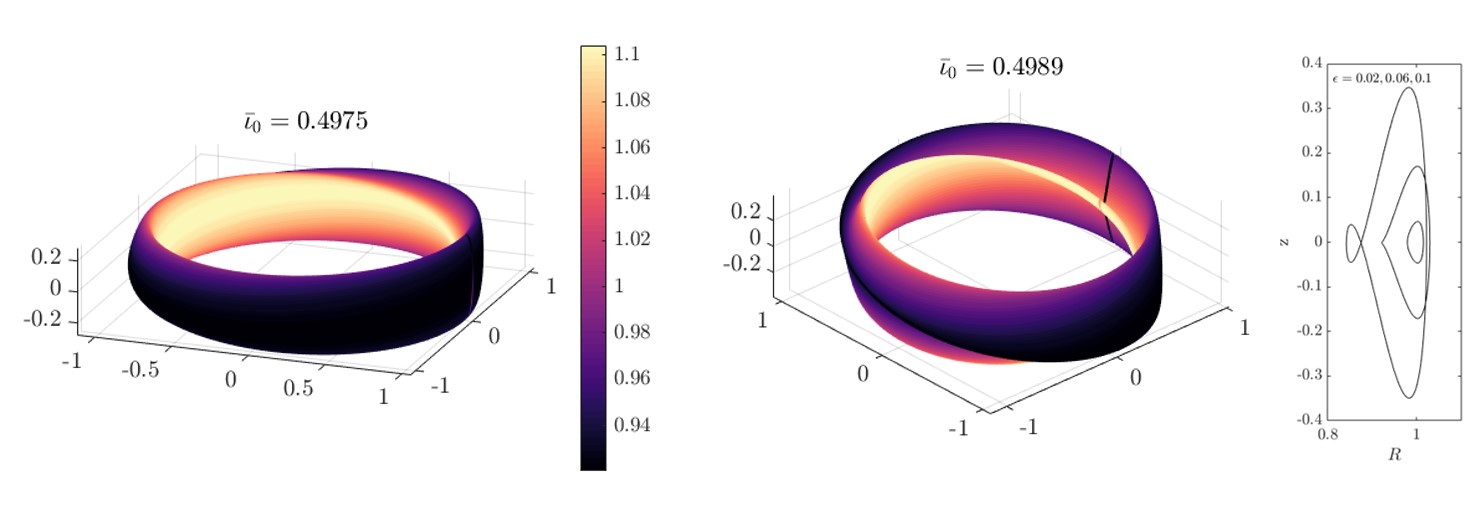}
     \caption{\textbf{Flux surfaces approaching the $\Bar{\iota}_0=1/2$ resonance.} The figures show the magnetic field magnitude on the magnetic flux surfaces at $\epsilon=0.1$ for two values of the rotational transform as the $\Bar{\iota}_0=1/2$ resonance is approached. The flux surface cross section corresponds to the figure on the right, and shows that flux surfaces start to overlap at lower values of $\epsilon$.}
     \label{fig:resSurf}
\end{figure*}

We show examples of the problems associated with choosing a rotational transform on axis that is close to the rational value of $\Bar{\iota}_0=1/2$ in Fig. \ref{fig:resSurf}. The enhanced shaping is evident, making surfaces overlap at lower values of $\epsilon$, and thus restricting the construction to larger-aspect-ratio stellarators. 


\section{Isotropic limit}

In the previous section, we have presented an explicit numerical example of a circular-axis quasisymmetric field through second order in the near-axis expansion. In doing so, we appear to have overcome the overdetermination problem that forbids such a construction in the isotropic limit\cite{garrenboozer1991b}. Now, it is natural to ask: can the procedure just employed yield isotropic solutions in some limit? This limit is expected to shed some light on the role played by the various free parameters (including $\bar{\Delta}_{20}$) in the construction of solutions. \par
To answer the question, it is first necessary to establish what, in the present context, is meant by the isotropic limit. For the solution to be \textit{isotropic}, the function $\Delta=(p_\parallel-p_\perp)/B^2$ must vanish. And it should do so systematically to all orders considered, that is, up to second order in the present context. \par
Doing so for the leading order, represented by the anisotropy on axis, is straightforward. One may simply set $\Delta_0=0$ as this function is an input for the construction. Setting $\Delta_0=0$ guarantees $\Delta_1=0$ (see Part I). However, the same is not true for the anisotropy at the second order. Though $\Delta_{22}=0$, $\Delta_{20}$ is generally not, given that it depends on the parameter $\bar{\Delta_{20}}$. \par
\begin{figure}[]
    \centering
    \includegraphics[width=0.45\textwidth]{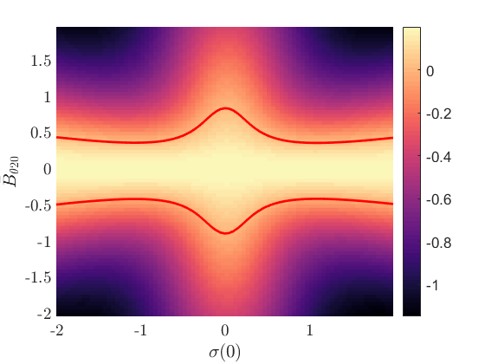}
     \caption{\textbf{Optimisation space for valid isotropic solutions.} The figure shows the space spanned by parameters $\Bar{B}_{\theta20}$ and $\sigma(0)$, with the color map corresponding to the deviation of $B_{\psi0}$ from its correct single-valued behaviour. The red line represents the loci of valid isotropic solutions. The field coefficients are those in the example before, but looking for isotropic solutions.}
     \label{fig:paramSpcIsotrop}
\end{figure}

For a truly isotropic solution then, $\Bar{\Delta}_{20}$ must somehow be reduced to zero. However, in the construction discussed above, this parameter serves as an output rather than an input.  Thus to construct an isotropic solution, it will be necessary to vary other available free parameters until an isotropic solution is found. Figure \ref{fig:paramSpcS0D0} is an example showing how $\bar{\Delta}_{20}$ changes with other parameters, in this case $\bar{\iota}_0$. For this example, there is a unique solution with $\bar{\Delta}_{20}=0$ in this space (ignoring the sign duplicity). Varying some other parameters such as $\sigma(0)$, the space of isotropic solutions may be shown to be larger, as shown in Fig. \ref{fig:paramSpcIsotrop}. 
\par
It is thus evident that there exists a whole set of parameter combinations that allow for a construction of isotropic quasisymmetric fields through second order. Upon first glance, this seems to be in stark contrast with the observation\cite{garrenboozer1991b} that the problem is overconstrained at second order and thus cannot be solved. However, further reflection suggests that this  contradiction is only an apparent one and stems from analysing the solvability of the expansion procedure solely by counting the number of equations and unknowns. In fact, our work suggests as a resolution that certain \textit{combinations of parameters effectively reduce the number of independent constraint equations allowing for a solution to be found}. \par
To make this observation stronger, we investigate the isotropic limit of the self-consistent set of equations $\Tilde{\mathrm{II}}_\mathrm{SC}^3$ in (\ref{eq:circLoopIIa}) and (\ref{eq:circLoopIIb}). In the `isotropic' limit, generally, the equations simplify quite significantly (considering $B_{\psi0}'\rightarrow0$ and many functions such as $\sigma$ or $B_{\theta20}$ to be constant, yet keeping a finite $\bar{\Delta}_{20}$),
\begin{align*}
    \frac{8B_0}{3\eta}\Bar{\iota}_0^2Y_{20}&\approx-B_0^2\tilde{p}_{31}^SB_{\alpha0}, \\
    \frac{16B_0}{3\eta}\Bar{\iota}_0^2\sigma Y_{20}&\approx B_0^2B_{\alpha0}\tilde{p}_{31}^C+B_{\alpha0}B_0\eta\Bar{\Delta}_{20}.
\end{align*} 
Upon first glance, the general structure of these equations might suggest that the problem is overdetermined. However, provided that
\begin{equation}
    \Bar{\Delta}_{20}=\frac{\Bar{\iota}_0}{\eta B_{\alpha0}}\left(2\sigma\Tilde{B}_{\psi11}^C-\Tilde{B}_{\psi11}^S\right)+2\frac{B_{\alpha1}}{B_{\alpha0}}, \label{eq:isoEqn}
\end{equation}
these equations will actually not be independent. In the isotropic situation, the left-hand-side is fixed to a value of 0, and thus Eq. (\ref{eq:isoEqn}) prescribes allowed combinations of coefficients. That when satisfying an equation like (\ref{eq:isoEqn}) the construction avoids overdetermination suggests that there could be particular coefficient combinations that would avoid overdetermination not only for more complex magnetic axis shapes, but also at higher order. This hypothesis could open the door to more `global' quasisymmetric solutions than those obtianed in present approaches such as [\onlinecite{landreman2019}]. We leave these explorations to future work. \par
Before concluding this section, we remark that Eqn. (\ref{eq:isoEqn}) also provides some insight into why solving for $\bar{\Delta}_{20}$ seems to always work in yielding a solution. A similar equation will hold approximately in moderately anisotropic systems, and a similar equation could be expected to exist in more general cases. With this in mind, one sees that an appropriate value $\bar{\Delta}_{20}$ equal to the right-hand-side may always be chosen for any given choice of the remaining free parameters. In a sense, one may think of this $\bar{\Delta}_{20}$ piece as being used to balance any left-over piece unable to balance with the isotropic $\n{j}\times\n{B}=\nabla p$ force.  


\section{Conclusion}
In this paper, we present an application of the near-axis expansion construction for a quasisymmetric magnetic field in anisotropic pressure force balance with a circular axis. A step-by-step procedure is given following Part I to explicitly construct a solution through second order in the distance from the axis. This is in contrast with the construction of Garren and Boozer\cite{garrenboozer1991b} which appears to lead to the problem of overdetermination at second order and beyond. \par
It is shown here that the isotropic limit may be taken quite naturally without falling into the problem of overdetermination. One is forced into particular combinations of parameters describing the field when doing so, which effectively modifies the number of independent constraint equations in the problem, circumventing [\onlinecite{garrenboozer1991b}]. \par
The anisotropic formulation presented here thus seems to offer a natural way into constructing quasisymmetric fields to higher orders. The avoidance of overdetermination suggests that there might exist a way of obtaining solutions in a more global sense, which in special case might be isotropic. A more definitive proof is left for future work.

\section*{Appendix A. Summary of Definitions}
Let the magnetic field be written,
\begin{align*}
    \n{B}=&B_\theta\nabla\chi+(B_\alpha-\bar{\iota}B_\theta)\nabla\phi+B_\psi\nabla\psi\\
    =&\nabla\psi\times\nabla\chi+\bar{\iota}\nabla\phi\times\nabla\psi,
\end{align*}
where $\bar{\iota}=\iota-N$, $\iota$ is the rotational transform and $N$ is the helicity of the magnetic field contours. The helical coordinate $\chi=\theta-N\phi$ is based on the set of generalised Boozer coordinates $\{\psi,\theta,\phi\}$. \par
Magnetic flux surfaces are described by the position vector,
\begin{equation*}
    \n{x}-\n{r}_0=X\hat{\kappa}+Y\hat{\tau}+Z\hat{b},
\end{equation*} 
where $\{X,~Y,~Z\}$ are space functions of the generalised Boozer coordinates, $\{\hat{b},\hat{\kappa},\hat{\tau}\}$ form the Frenet set of unit vectors associated with the magnetic axis, and $\n{r}_0$ represents the axis itself. \par
Concerning the functions pertaining to force balance, perpendicular and parallel components of the pressure are defined ($p_\perp$ and $p_\parallel$ respectively). An anisotropy function is also defined $\Delta=(p_\parallel-p_\perp)/B^2$, which by the physical requirement of positivity of the pressure must satisfy the inequality $\Delta>-p_\perp/B^2$.\par
All of these functions are to be expanded, as part of the near-axis expansion, in Taylor-Fourier series such as,
\begin{equation*}
    B_\theta=\sum_{n=0}^\infty\epsilon^n{\sum_{m=0|1}^{n}}\left[B_{\theta nm}^c(\phi)\cos m\chi+B_{\theta nm}^s(\phi)\sin m\chi\right].
\end{equation*} 
Some functions with a particular form of symmetry such as the magnetic field, have simplified forms of the expansion with constant coefficients,
\begin{equation*}
    \frac{1}{B^2}=B_0+\sum_{n=1}^\infty\epsilon^n \sum_{m=0|1}^n\left(B_{nm}^c\cos m\chi +B_{nm}^s\sin m\chi\right).
\end{equation*}
Finally, flux functions such as $B_\alpha$ or $\bar{\iota}$ will have standard Taylor expansions,
\begin{equation*}
    \iota(\psi)=\sum_{n=0}^\infty\epsilon^{2n}\iota_n.
\end{equation*}
For more details we refer the reader to Part I.

\section*{Appendix B. Explicit constructions through third order for \textit{looped} II$^3$}
Let us start by writing II$^3$ explicitly. From Eq. (65) of Part I, we write
\begin{align*}
    B_0&[(1-\Delta_0)B_{\theta3,1}^C]'-B_0\Bar{\iota}_0(\Delta_0-1)B_{\theta3,1}^S=B_0^2B_{\alpha0}p_{3,1}^S-\\
    &-\frac{B_{\alpha0}}{4}\left[2B_{11}^S\Delta_0-8B_0B_{11}^Cp_{2,2}^S+2B_{2,2}^S(2B_0p_{11}^C+\right.\\
    &\left.+\Delta_{11}^C)-2B_{22}^C\Delta_{11}^S-B_{11}^C\Delta_{2,2}^S\right]+B_{11}^C(2B_{\theta20}\Delta_0'+\\
    &+(\Delta_0-1)B_{\theta20})-B_0(-\Delta_{11}^CB_{\theta20}'+B_0B_{\theta20}p_{11}^C{}')\\
    B_0&[(1-\Delta_0)B_{\theta3,1}^S]'+B_0\Bar{\iota}_0(\Delta_0-1)B_{\theta3,1}^C=-B_0^2B_{\alpha0}p_{3,1}^C-\\
    &-\frac{B_{\alpha0}}{4}\left[2B_{11}^C\Bar{\iota}_0\Delta_0B_{\theta20}-4B_0^2\Bar{\iota}_0B_{\theta20}p_{11}^C+\right.\\
    &\left.+B_{\alpha1}(-2B_{11}^C\Delta_0+4B_0^2p_{11}^C)+B_{\alpha0}(-2B_{3,1}^C\Delta_0+\right.\\
    &\left.+(B_{11}^C)^2p_{11}^C+8B_0B_{20}^Cp_{11}^C-4B_0B_{22}^Cp_{11}^C-2B_{22}^C\Delta_{11}^C+\right.\\
    &\left.+B_{11}^C(8B_0p_{22}^C-2\Delta_{20}+\Delta_{22}^C)-2B_{22}^S\Delta_{11}^S)-\right.\\
    &\left.-4B_0\Delta_{11}^SB_{\theta20}'+4B_0^2B_{\theta20}p_{11}^S{}'\right].
\end{align*}
These clearly look like ordinary differential equations in $B_{\theta31}$, but they depend on the third-order form of the pressure as well. Thus, we should be careful and write down the expression for $p_3$ using III$^1$ to eliminate it (see Eq. (66) in Part I). This is algebraically convoluted, but important. We write
\begin{align*}
    p_{3,1}^C&=\frac{B_{\alpha0}}{12B_0}\left[2B_{\alpha1}(B_{11}^C(-4+5\Delta_0)-2B_0(B_0p_{11}^C-\right.\\
    &\left.-2\Delta_{11}^C))+B_{\alpha0}(6B_{31}^C\Delta_0-3(B_{11}^C)^2p_{11}^C-8B_{20}^Cp_{11}^C-\right.\\
    &\left.-4B_0B_{22}^Cp_{11}^C+4B_{20}^C\Delta_{11}^C+2B_{22}^C\Delta_{11}^C+B_{11}^C(-16B_0p_{20}-\right.\\
    &\left.-8B_0p_{22}^C+\Delta_{20}+\Delta_{22}^C)-8(B_0\Bar{\iota}_0((\Delta_0-1)B_{\psi11}^S+\right.\\
    &\left.+B_{\psi0}\Delta_{11}^S)+B_{11}^C((\Delta_0-1)B_{\psi0}'+B_{\psi0}\Delta_0')+\right.\\
    &\left.+B_0(\Delta_{11}^CB_{\psi0}'+B_{\psi11}^C\Delta_o'-B_{\psi11}^C{}'+\Delta_0B_{\psi11}^C{}'+\right.\\
    &\left.+B_{\psi0}\Delta_{11}^C{}'))\right] 
\end{align*}
\begin{align*}
    p_{3,1}^S&=\frac{B_{\alpha0}}{12B_0}\left[B_{\alpha0}(6B_{31}^S\Delta_0-8B_0B_{11}^Cp_{22}^S+B_{22}^S(-4B_0p_{11}^C+\right.\\
    &\left.+2\Delta_{11}^C)+4B_{20}\Delta_{11}^S-2B_{22}^C\Delta_{11}^S+B_{11}^C\Delta_{22}^S)+\right.\\
    &\left.+8B_0(\Bar{\iota}_0((\Delta_0-1)B_{\psi11}^C+B_{\psi0}\Delta_{11}^C)+B_{\alpha1}\Delta_{11}^S-\right.\\
    &\left.-\Delta_{11}^SB_{\psi0}'-B_{\psi11}^S\Delta_0'+B_{\psi11}^S{}'-\Delta_0B_{\psi11}^S{}'-B_{\psi0}\Delta_{11}^S{}')\right].
\end{align*}
Finally, and because these depend on $B_{\psi1}$, expressions for these are needed to complete the \textit{loop} eeqautions. So, using $C_\perp^2$, we obtain
\begin{align*}
    B_{\psi11}^S=&\frac{3}{2}B_{\theta31}^C-\frac{8\Bar{\iota}_0\sigma}{\eta}Y_{20}+\frac{B_0}{\eta^3}\left[\Bar{\iota}_0\left(-\frac{\eta^4}{B_0}+12(\sigma^2+\right.\right.\\
    &\left.\left.+1)\right)+8\sigma'\right]B_{\psi0}'-\frac{4}{\eta}Y_{20}'+\Tilde{B}_{\psi11}^S \\
    B_{\psi11}^C=&-\frac{3}{2}B_{\theta31}^S-\frac{4\Bar{\iota}_0}{\eta}Y_{20}-\left(\eta+\frac{4B_0}{\eta^3}(1+\sigma^2)\right)B_{\psi0}''+\\
    &+\Tilde{B}_{\psi11}^C,
\end{align*}
where
\begin{align*}
    \Tilde{B}_{\psi11}^C&=-\frac{4\Bar{\iota}_0}{\eta}(\Tilde{Y}_{22}^C-\sigma\Tilde{Y}_{22}^S)+\frac{2}{\eta}B_{\theta20}\left[\sigma(\Tilde{X}_{20}-X_{22}^C)+X_{22}^S\right]+\\
    &+\frac{2\eta}{B_0}\left[\Tilde{Y}_{22}^S(\Tilde{Z}_{20}-Z_{22}^C)+\Tilde{Y}_{22}^CZ_{22}^S\right]+\frac{2\sigma}{\eta}\Tilde{Y}_{22}^C{}'\\
    \Tilde{B}_{\psi11}^S&=-\frac{4\Bar{\iota}_0}{\eta}(\Tilde{Y}_{22}^S+\sigma\Tilde{Y}_{22}^C)-\frac{2\eta}{B_0}\Tilde{Y}_{22}^C\Tilde{Z}_{20}-\frac{B_{\theta20}}{2\eta}\left[3\eta^2-\right.\\
    &\left.-4\tilde{X}_{20}-4X_{22}^C+4\sigma X_{22}^S\right]-\frac{2}{\eta}(\Tilde{Y}_{22}^C{}'-\sigma\tilde{Y}_{22}^S{}')+\\
    &+\frac{B_{\alpha0}}{\eta}\left[\eta^2\Tilde{Y}_{22}^SB_{\theta20}-2\eta^2\Bar{\iota}_0X_{22}^C+4\Tilde{Z}_{20}\sigma X_{22}^C+\right.\\
    &\left.+4\Tilde{Z}_{20}X_{22}^S-4\sigma\Tilde{X}_{20}Z_{22}^C+4X_{22}^SZ_{22}^C+\eta^2Z_{22}^S-\right.\\
    &\left.-4\Tilde{X}_{20}Z_{22}^S-4X_{22}^CZ_{22}^S+\eta^2X_{22}^S{}'\right].
\end{align*}
Plugging explicitly the form for $B_{\psi11}$ into the expressions for the pressure $p_{31}$, and $p_{31}$ into the equations for $B_{\theta31}$, we obtain
\begin{align}
    &\left\{\frac{2}{3}B_0(\Delta_{11}^S-\Bar{\iota}_0\eta\Delta_0')+\frac{8B_0^2}{\eta^3}\Bar{\iota}_0\left[(1+\sigma^2)(\Delta_0-1)\right]'+\right.\nonumber\\
    &\left.+\frac{16B_0^2}{3\eta^3}\left[(\Delta_0-1)\sigma'\right]'\right\}B_{\psi0}'+\nonumber\\
    &\left\{\frac{32B_0^2}{3\eta^3}\Bar{\iota}_0(\Delta_0-1)(1+\sigma^2)+\frac{16B_0^2}{3\eta^3}(\Delta_0-1)\sigma'\right\}B_{\psi0}''+\nonumber\\
    &\left\{\frac{8B_0}{3\eta}\Bar{\iota}_0^2(\Delta_0-1)+\frac{16B_0}{3\eta}\Bar{\iota}_0\left[(1-\Delta_0)\sigma\right]'\right\}Y_{20}+\nonumber\\
    &\left\{\frac{16B_0}{3\eta}\Bar{\iota}_0\sigma(a-\Delta_0)-\frac{8B_0}{3\eta}\Delta_0'\right\}Y_{20}'+\nonumber\\
    &\left\{\frac{8B_0}{3\eta}(1-\Delta_0)\right\}Y_{20}''-\nonumber\\
    &-B_0^2B_{\alpha0}\Tilde{p}_{31}^S+\frac{B_{\alpha0}}{2}\left[B_{31}^S\Delta_0+8B_0^2\eta p_{22}^S+B_{22}^S(-2\eta\Delta_0+\right.\nonumber\\
    &\left.+\Delta_{11}^C)-B_{22}^C\Delta_{11}^S+B_0\eta\Delta_{22}^S\right]+B_0\left[3\eta B_{\theta20}\Delta_0'+\right.\nonumber\\
    &\left.+B_{\theta20}'(2\eta(\Delta_0-1)-\Delta_{11}^C)\right]=0, \label{eq:circLoopIIafull}
\end{align}
and
\begin{align}
    &\left\{\frac{2}{3}B_0\eta(1-\Delta_0)(1-\Bar{\iota}_0^2)-\frac{2}{3}B_0\Delta_{11}^C+\frac{8B_0^2}{\eta^3}\Bar{\iota}_0^2(1-\Delta_0)\times\right.\nonumber\\
    &\left.\times(1+\sigma^2)+\frac{16B_0^2}{3\eta^3}\Bar{\iota}_0(1-\Delta_0)\sigma'\right\}B_{\psi0}'+\nonumber\\
    &\left\{\frac{8B_0^2}{3\eta^3}\left[(1+\sigma^2+\frac{\eta^4}{4B_0})(\Delta_0-1)\right]'\right\}B_{\psi0}''+\nonumber\\
    &\left\{\frac{8B_0^2}{3\eta^3}(\Delta_0-1)\left(1+\sigma^2+\frac{\eta^4}{4B_0}\right)\right\}B_{\psi0}'''+\nonumber\\
    &\left\{\frac{8B_0}{3\eta}\Bar{\iota}_0\left[2\Bar{\iota}_0(\Delta_0-1)\sigma+\Delta_0'\right]\right\}Y_{20}+\nonumber\\
    &\left\{\frac{16B_0}{3\eta}\Bar{\iota}_0(\Delta_0-1)\right\}Y_{20}'+\nonumber\\
    &B_0^2B_{\alpha0}\Tilde{p}_{31}^C-\frac{B_{\alpha0}}{2}\left[-2B_0\eta\Tilde{\Delta}_{20}^C+2B_0^2\eta\Tilde{p}_{20}+2B_0\eta^3\Delta_0+\right.\nonumber\\
    &\left.+4\eta B_{20}\Delta_0-2\eta B_{22}^C\Delta_0+B_{31}^C\Delta_0+8B_0^2\eta p_{22}^C+\right.\nonumber\\
    &\left.+B_{22}^C\Delta_{11}^C+B_0\eta\Delta_{22}^C+B_{22}^S\Delta_{11}^S\right]-B_0\Delta_{11}^SB_{\theta20}'=0,\label{eq:circLoopIIbfull}
\end{align}
where
\begin{align*}
    \Tilde{p}_{31}^C&=-\frac{\eta}{3B_0}\Tilde{\Delta}_{20}+3\eta\Tilde{p}_{20}-\frac{2\Bar{\iota}_0}{3B_0B_{\alpha0}}(\Delta_0-1)\Tilde{B}_{\psi11}^S+\\
    &+\frac{\eta^3}{B_0}\Delta_0+\frac{2\eta B_{20}}{3B_0^2}\Delta_0+\frac{\eta B_{22}^C}{3B_0^2}\Delta_0+\frac{4}{3}\eta p_{22}^C-\\
    &-\frac{2B_{\alpha1}}{3B_0B_{\alpha0}}(-2\eta+2\eta\Delta_0-\Delta_{11}^C)+\frac{B_{20}}{3B_0^2}\Delta_{11}^C+\\
    &+\frac{B_{22}^C\Delta_{11}^C}{6B_0^2}-\frac{\eta \Delta_{22}^C}{6B_0}+\frac{B_{22}^S\Delta_{11}^S}{6B_0^2}+\\
    &+\frac{2}{3B_0B_{\alpha0}}\left[\Tilde{B}_{\psi11}^C(1-\Delta_0)\right]',
\end{align*}  
and
\begin{align*}
    \Tilde{p}_{31}^S&=\frac{2\Bar{\iota}_0}{3B_0B_{\alpha0}}(\Delta_0-1)\Tilde{B}_{\psi11}^C+\frac{2B_{\alpha1}}{3B_0B_{\alpha0}}\Delta_{11}^S+\\
    &+\frac{1}{6B_0^2}\left[3B_{31}^S\Delta_0+8B_0^2\eta p_{22}^S+B_{22}^S(2\eta \Delta_0+\Delta_{11}^C)+\right.\\
    &\left.+2B_{20}\Delta_{11}^S-B_{22}^C\Delta_{11}^S-B_0\eta\Delta_{22}^S\right]+\\
    &+\frac{2}{3B_0B_{\alpha0}}\left[\Tilde{B}_{\psi11}^S(1-\Delta_0)\right]'.
\end{align*}
In the main text this equation is presented schematically, labeling the factors associated to each one of the terms.

\section*{Acknowledgements}
 This research is primarily supported by a grant from the Simons Foundation/SFARI (560651, AB). 

\section*{Data availability}
The data that support the findings of this study are available from the corresponding author upon reasonable request.

\bibliography{avoidOverGB}

\end{document}